\documentclass[prb,twocolumn,nofootinbib]{revtex4}
\usepackage{graphicx}
\usepackage{amssymb}

\usepackage{amsmath,amssymb,amsfonts,times,graphicx}
\usepackage[ps2pdf,bookmarks=true,colorlinks,linkcolor=red,urlcolor=blue,citecolor=blue]{hyperref}
\usepackage[usenames]{color}
\usepackage{dcolumn}

\begin{document}
\title{Entanglement Structure of Deconfined Quantum Critical Points}
\author{Brian Swingle}
\affiliation{Department of Physics, Massachusetts Institute of Technology, Cambridge, MA 02139}\affiliation{ Department of Physics, Harvard University, Cambridge, MA 02138}
\author{T. Senthil}
\affiliation{Department of Physics, Massachusetts Institute of Technology, Cambridge, MA 02139}
\begin{abstract}
We study the entanglement properties of deconfined quantum critical points.  We show not only that these critical points may be distinguished by their entanglement structure but also that they are in general more highly entangled that conventional critical points.  We primarily focus on computations of the entanglement entropy of deconfined critical points in $2+1$ dimensions, drawing connections to topological entanglement entropy and a recent conjecture on the monotonicity under RG flow of universal terms in the entanglement entropy.  We also consider in some detail a variety of issues surrounding the extraction of universal terms in the entanglement entropy.  Finally, we compare some of our results to recent numerical simulations.
\end{abstract}
\maketitle

\section{Introduction}
Conventional ground states of bosonic quantum matter are distinguished by the presence or lack thereof of broken symmetry and associated long range order (LRO). These ground states have prototypical wavefunctions that are direct products of local degrees of freedom, and hence contain only short range quantum entanglement. In the last several years much attention has been lavished on ordered states which do not fit the broken symmetry/long range order paradigm. The best studied and simplest examples are gapped topological phases with a low energy effective description in terms of a topological field theory as introduced in Refs. \onlinecite{toporder1,toporder2}.  More complex examples also exist in the form of gapless liquid-like phases which also do not fit the broken symmetry paradigm. A common feature of all these `exotic' states is that their prototypical ground state wavefunctions are not direct products of any local degrees of freedom. These states have long range entanglement (LRE) in their ground state wavefunction. Thus the study of the entanglement properties of the ground state of various quantum phases and their associated phase transitions has become a topic of fundamental importance. Entanglement based ideas have led to a classification of gapped quantum phases in one dimension described in Refs. \onlinecite{mps_classify1,mps_classify2}, efficient ways to characterize critical theories in one dimension described in Refs. \onlinecite{vidal_mera,vidal}, a partial characterization of topological phases in two dimensions as shown in Refs. \onlinecite{topent1,topent2}, and much more.

Due to these and other considerations, there has been a recent profitable exchange of ideas between quantum information science and quantum many-body physics. Broadly speaking, the study of entanglement entropy has provided us with a rough guide to the landscape of many-body states.  It guides us to the physical region of Hilbert space where many-body states live.  Most importantly, entanglement entropy is at present one of the best available quantities to concretely probe the structure of long range entanglement in quantum many-body ground states. In this paper, as part of the broader effort to understand the entanglement properties of zero temperature phases and phase transitions, we describe the structure of entanglement at certain deconfined quantum critical points in higher dimensions with a special focus on two dimensions.  In particular, we will compute universal subleading corrections to the entanglement entropy for a variety of deconfined quantum critical points in two dimensions described in Refs. \onlinecite{deccp,deccplong}.

Deconfined critical points lie outside the scope of the Landau paradigm for phase transitions because they are most naturally described in terms of ``deconfined" degrees of freedom that exist at the critical point (but perhaps not in either phase).  The emergence of these deconfined degrees of freedom and associated gauge fields suggests that the ground state of these critical points may have ``stronger" long range entanglement as compared to conventional quantum critical points ({\em i.e} those described within the Landau paradigm). It is therefore interesting to explore their entanglement properties in some detail. Deconfined critical points are useful paradigmatic examples of exotic gapless systems with strongly non-quasiparticle like excitation spectra.  Studying their entanglement structure is a step in exploring the relationship between the breakdown of the concept of a quasiparticle to describe the excitations of a many-body system and possible long range entanglement in the corresponding ground state wavefunction.

A simple example of a deconfined critical point is provided by a quantum phase transtion between a superfluid and a $Z_2$ topological Mott insulator of bosons at integer or half-integer filling on a lattice (see Refs. \onlinecite{z2long,SeMo02,MoSe02,GrSe10,melko1,melko2}). The latter phase has gapped fractional charge-$1/2$ bosons (dubbed chargons) and gapped $Z_2$ vortices (dubbed visons) which are mutual semions. The transition to the superfluid goes hand in hand with the destruction of topological order. Despite the presence of a natural Landau order parameter for the superfluid phase, the critical theory is not described by a conventional Landau-Ginzburg-Wilson action in terms of this order parameter. Rather it is most simply described as a condensation transition of the fractionally charged boson in the $D = 2+1$ $XY$ universality class.  The exponents $\nu$ and $z$ are the same as in the usual $XY$ universality class, but the anomalous dimension of the superfluid order parameter is large (effectively because the superfluid order parameter is a composite field).  The universality class of this transition has been denoted $XY^*$ to distinguish it from $XY$. Coming from the superfluid side the deconfinement of the fractional boson occurs already at the critical point and continues into the Mott phase. This deconfined critical point provides an excellent setting for us to contrast the entanglement behavior with conventional quantum critical points by simply comparing the universal subleading behavior of the entanglement entropy of the $XY^*$ and $XY$ transitions. Indeed, we will show that this universal part is different for the two transitions with the difference precisely due to the emergent topological structure of the proximate Mott insulator.

Another well studied example of a deconfined critical point is that of a continuous Landau-forbidden quantum phase transition between Neel and Valence Bond Solid (VBS) phases of spin-$1/2$ quantum antiferromagnets on a square lattice (see Refs. \onlinecite{deccp,deccplong}). The critical field theory is naturally described in terms of emergent gapless bosons that carry fractional spin and are coupled to an emergent gapless $U(1)$ gauge field. In this case we argue that the universal part should be correctly given by calculations in the non-compact $CP^1$ conformal field theory that describes the low energy physics - microscopic lattice models differ from this field theory only by irrelevant operators which do not affect the universal contribution to the entanglement entropy.

Further interest in these particular models comes from pioneering numerical work reported in Refs. \onlinecite{qmc_renyi1,melko0}.  They studied the entanglement properties of a simple realization of an $XY^*$ transition.  We will analyze their calculations in the context of our results later in the paper.  Even more recently, Ref. \onlinecite{ee_anom_sq_heis} studied various universal terms in the entanglement entropy of the square lattice Heisenberg model.  That work raised certain puzzles associated, for example, with corners.  Our work here suggests one possible resolution of those puzzles.  They compare their Monte Carlo and series results for corners with that of two free bosons (a non-compact spin wave description).  However, it may be important to remember the compactness of the actual order parameter.  In the case of an $XY$ ordered phase, the compactness of the order parameter may be approached by coupling a non-compact free boson to a $Z$ gauge field.  As we will argue repeatedly in this work, such couplings typically have the effect of shifting universal terms.  Our work below leads us to believe that the gauge field coupling would make the corner contribution more negative in closer agreement with the numerical results.  At the very least, there could be a crossover length scale that would be have to be exceeded to truly detect the non-compact boson corner contribution.  Simply speaking, we are reducing the number of degrees of freedom and hence should reduce the entanglement.  We also note that the $J-Q$ model introduced by Sandvik in Ref. \onlinecite{JQsandvik} seems to realize the deconfined Neel to VBS transition, and since this model is sign problem free, it is also a very attractive target for investigations of entanglement entropy at deconfined critical points.

Now we outline the details of our calculations.  The entanglement entropy of a region $R$ is defined as $S(R) = - \mbox{tr}_R (\rho_R \ln{\rho_R})$ where $\rho_R$ is the density matrix for $R$. When considering multiple regions, we simply concatenate the labels, hence $S(R_1 R_2)$ is the entropy of the combined region $R_1 \cup R_2$.  The basic rule governing the entanglement entropy of a local many-body system is the boundary law.  In $d$ spatial dimensions the entanglement entropy of a region $R$ of linear size $L$ is roughly $L^{d-1}$ i.e. it is proportional to the size of the boundary of the region as reviewed in Ref. \onlinecite{arealaw1}.  Exceptions to this rule include gapless systems in one dimension as shown in Refs. \onlinecite{geo_ent,eeqft} and systems with a Fermi surface in higher dimensions as argued in Refs. \onlinecite{fermion1,fermion2,bgs_ferm1,ee_proj_fs}, but the rule is obeyed by a wide variety of gapped and gapless phases in higher dimensions.  It is a super-universal feature of local many-body systems.  That being said, the prefactor of the boundary law term in the entanglement entropy is not directly meaningful because it contains details of the microscopic regulator.  Different microscopic realizations of the same phase or phase transition can have different values for this prefactor.  In other words, while the existence of the boundary law term is super-universal, its precise value is not universal at all.

All this information can be conveniently summarized by an intuitive scaling argument described, for example, in Ref. \onlinecite{qft_mi}.  To obtain the entanglement entropy of a region $R$ we take a renormalization group perspective and add up the contributions to the entanglement entropy from each scale.  The key assumptions are locality in space and approximate locality in scale (so that we may add contributions from different scales).  Let $r$ indicate the length scale of interest in our many-body system.  $r$ runs from the ultraviolet cutoff $\epsilon$ (the lattice scale) to $\infty$ in the long wavelength limit.  Spatial locality suggests that the contribution to the entanglement entropy at scale $r$ is proportional to the size of the boundary measured in units of $r$: $dS(r) \sim (L/r)^{d-1}$.  We integrate this contribution from the cutoff $\epsilon$ to the minimum of the the correlation length $\xi$ (gapped) and the region size $L$ (gapless).  The appropriate renormalization group measure is $d \ln{r} = dr/r$.  The result of this simple calculation is a non-universal boundary law for gapped phases in $d=1$, a universal logarithmic dependence for gapless phases in $d=1$, and a non-universal boundary law for gapped and gapless phases in $d>1$.  Thus as we saw above, to extract universal features in higher dimensions we must turn to sub-leading terms in the entanglement entropy.

The general structure of sub-leading terms can be understood via a generalization of the scaling argument given above.  In Ref. \onlinecite{qft_mi} a ``twist field" formalism was used to investigate the short distance structure of entanglement.  The result of this expansion is that only even or odd terms in an $L/r$ expansion appear as appropriate; furthermore, the pattern obtained in Ref. \onlinecite{qft_mi} agrees with known results (see Ref. \onlinecite{holo_ee_origin,casini_disc_ee}) in $1 \geq d \leq 3$ spatial dimensions.  In two dimensions, the entanglement entropy of a smooth region of linear size $L$ in a gapped phase has the form $S(L) = \alpha L - \beta + ...$.  The coefficient $\alpha$ is non-universal, however, the coefficient $\beta$ is universal and partially characterizes the topological order of the phase as shown in Refs. \onlinecite{topent1,topent2}.  $\beta = 0$ for any phase adiabatically connected to a trivial product state e.g. a bosonic Mott insulator at integer filling.  For topological phases characterized by a modular S-matrix $\mathcal{S}^a_b$, we have $\beta = \ln{\mathcal{S}^0_0} = - \ln{ \mathcal{D}}$ with $\mathcal{D}$ the total quantum dimension, a result we will re-derive later.  In a certain precise sense the coefficient $\beta$ increases with the quantum complexity of the topological phase.  Gapless phases in two dimensions have a similar form for their entanglement entropy, but other terms are also possible, for example, corners will give logarithmic terms in the entanglement entropy (see Ref. \onlinecite{eecorner} for a sample calculation and Ref. \onlinecite{qft_mi} for a scaling argument).  Note that there is some subtlety in giving meaning to $\beta$ in the presence of such logarithmic corner terms.  Furthermore, in a gapless phase $\beta$ will depend on region shape, a fact we will sometimes emphasize by writing $\beta = \beta(R)$.  For example, a long thin strip will have $\beta(L,w) \sim L/w$ with $L$ the length and $w$ the width of the strip.  It is thus important to give a precise protocol for defining and extracting $\beta$.  We will return to these questions below.

Before moving on, we wish to comment on a slightly counter-intuitive feature of the above discussion.  We claimed that phases with larger $\beta$ are more entangled, but looking at the formulae above, it seems that a larger $\beta$ corresponds to a smaller entanglement entropy.  We may understand this as follows.  For any large smooth region in a gapped phase, the entropy has the form $S = \alpha L - \beta$.  We may change the value of $\alpha$ by adiabatically deforming the Hamiltonian within the phase, but the value of $\beta$ is fixed and universal.  Thus because $\beta$ contributes negatively to a positive quantity, the entanglement entropy, we know that in any phase with non-zero $\beta$, we cannot reduce $\alpha$ to zero by deforming the Hamiltonian.  In other words, a phase with non-zero $\beta$ has no microscopic Hamiltonian in which the ground state is a product state.  As $\beta$ increases, the ``obstruction" to a product state also increases, thus establishing that larger $\beta$ corresponds to a ``more entangled" phase of matter.

Our basic result for the $XY^*$  transition is that for simply connected regions $\beta_{XY^*} = \beta_{XY} + \beta_{Z_2}$ where $\beta_{Z_2} = \ln{2}$ is the topological entanglement entropy of a $Z_2$ gauge theory. As described above this simple result explicitly shows how the universal entanglement entropy acquires an additional topological contribution not present at the $XY$ critical point. This gives precise meaning to the notion that this deconfined critical point has stronger entanglement than the corresponding conventional critical point. In addition to the simple example of the $XY^*$ transition, we obtain similar results for an infinite class of similar deconfined critical points.  These deconfined critical points consist of bosons undergoing a superfluid-insulator transition where the insulator is an exotic fractionalized phase.  The description is in terms of emergent bosonic degrees charged under deconfined discrete gauge fields.  We also argue that the entanglement entropy of lattice models realizing the deconfined critical point separating Neel and VBS phases has the same universal properties as the critical $CP^1$ model.  We also explain why the conjecture of Ref. \onlinecite{highercurv_ee} concerning the existence of an entanglement c-theorem shows that the $CP^1$ model is more highly entangled than the usual $O(3)$ fixed point.

This paper is organized as follows.  First, we describe the $XY^*$ transition and compute the entanglement entropy directly from a model wavefunction.  Second, we generalize this argument to a large class of models using a partition function argument recently described in Ref. \onlinecite{geo_ent_edge_proof}.  Third, we discuss the properties of the $CP^1$ model and its field theory.  Fourth, we discuss a variety of procedures for defining and extracting $\beta$.  Finally, we conclude with a discussion of the broader relevance of our results.

\section{$XY^*$ transition}
\subsection{Description of transition}
Consider a system of bosons on a lattice at a commensurate filling of, say $1/2$, per site on average in two space dimensions described by a Hamiltonian with the general structure
\begin{equation}
H = -\sum_{ij} t_{ij} \left(\psi^\dagger_i \psi_j + h.c \right) + ...
\end{equation}
with $\psi_i^\dagger$ being the creation operator for a boson at site $i$.
For concreteness we will specialize to a square lattice though our results apply equally to other lattices. If the boson hopping dominates over all interaction terms a uniform superfluid phase will result. With increasing strength of short ranged repulsive interactions a variety of Mott insulating phases  result. Some simple possibilities are Mott insulators that break lattice symmetry such as checkerboard density wave or bond-centered stripe insulators. Our focus will however be on Mott insulators that preserve all symmetries of the Hamiltonian. For a square lattice and at half-filling
such Mott insulators are necessarily exotic (see Refs. \onlinecite{oshikawa,hastings}): they either have topological order and a gap to all excitations in the bulk or some even more exotic order and gapless excitations. We will mainly restrict attention to the simplest such possibility, namely a gapped Mott insulator with topological order that is described by the deconfined phase of a $Z_2$ gauge theory. Such a phase has gapped fractionalized excitations (dubbed chargons) that carry $1/2$ the boson charge. In addition here are gapped charge-neutral vison excitations that are mutual semions with the chargons. For various examples of microscopic boson models that realize such a phase and the associated phase transitions, see Refs. \onlinecite{BaFiGi02,SeMo02,MoSe02,melko0,melko1,melko2}.

Being topologically ordered the fractionalized Mott insulator has Long Range Entanglement in its ground state that is partially captured by the concept of Topological Entanglement Entropy. Recent numerical work has successfully demonstrated this property for a specific model. Here we are interested in the quantum phase transition between the superfluid state and the fractionalized Mott insulator as the interaction strength is varied at fixed filling. This transition is conveniently described as a condensation of the chargons while the vison gap stays non-zero. Specifically if we describe the chargons by a field $b$, the transition is described by the action of the space-time $D = 3$ dimensional $XY$ fixed point
$S_{XY}[b]$. Despite this however the transition is not quite in the $3D$ $XY$ universality class. This is because the field $b$ itself carries $Z_2$ gauge charge and hence is not directly observable. The physical boson operator $\psi \sim b^2$ and hence has a large anomalous dimension $\eta \approx 1.4$. Indeed the only allowed physical operators at the fixed point are the subset of the operators at the $3D$ $XY$ fixed point which are invariant under the local $Z_2$ gauge transformation associated with changing the sign of $b_i$ at any given site. The fixed point describing the superfluid- fractionalized Mott insulator transition is thus distinct from the $3D$ $XY$ fixed point and is denoted as an $XY^*$ fixed point. This \onlinecite{z2long,MoSe02,SeMo02,GrSe10,melko1,melko2} and other similar transitions \onlinecite{ChSeSa1,ChSeSa2,IsSeKi05,PhysRevB.81.144432} have been described in a number of contexts in previous papers. Most recently this transition has been explored numerically in Ref. \onlinecite{melko2} and the predicted large $\eta$ and other properties have been demonstrated.

As the simplest example of a non-Landau transition out of an ordered phase the $XY^*$ transition provides a good opportunity to examine the entanglement structure of an exotic phase transition. Our goal will be to compare the universal structure of the long range entanglement at the $XY^*$ fixed point with that of the $XY$ fixed point. This task is simplified by noting that as far as the universal properties of the transition are concerned we might as well take the vison gap to infinity. As the visons correspond to the magnetic flux of the gauge field this means that the `electric' field lines of the gauge field can be taken to be tensionless.

\subsection{Exact wavefunction computation}
Consider an $L\times L$ square lattice with bosons $b_i$ on the sites and $Z_2$ gauge fields $\sigma^z_{ij}$ on the links.  The bosons are charged under the gauge field and we take the gauge dynamics at the lattice scale to be tensionless.  The Hamiltonian is
\begin{eqnarray}
H = - w\sum_{<ij>} \sigma^z_{ij} b_i^+ b_j + \mbox{h.c.} - \mu \sum_i n_i \nonumber \\ + U \sum_i n_i (n_i -1) - K \sum_{ijkl \in \square } \sigma^z_{ij} \sigma^z_{jk} \sigma^z_{kl} \sigma^z_{li}
\end{eqnarray}
with $n_i = b_i^+ b_i $.  We also impose the constraints $\mathcal{C}_i = e^{i \pi n_i} \sigma^x_{i i+x} \sigma^x_{i i-x} \sigma^x_{i i+y} \sigma^x_{i i-y} = 1$. To study the $XY^*$ transition we tune the boson filling to be one per site and study the superfluid to insulator transition in the presence of the gauge field.
Because we are in the tensionless limit, the gauge fields $\sigma^z_{ij}$ are constants of the motion.  Choose a gauge where $\sigma^z_{ij} = 1$ for all $ij$ and call the ground state of the boson model $|B\rangle$.  Imposing the constraints attaches the ends of electric field lines to the bosons.

Write the boson state as $| B \rangle  = \sum_p c_p |B_p \rangle $ where $p$ is an index that records the even/odd parity of $n_i$ at each lattice site.  Thus $p$ runs over $2^{2L^2}$ values.  We take $\langle B_p | B_q \rangle = \delta_{p q}$.  The index $p$ records $e^{i \pi n_i} $ for each site and hence $\sigma^x_{i i+x} \sigma^x_{i i-x} \sigma^x_{i i+y} \sigma^x_{i i-y}$ for each site.  There are $2^{2L^2}$ states in the unconstrained gauge space.  For each configuration of $\sigma^x_{i i+x} \sigma^x_{i i-x} \sigma^x_{i i+y} \sigma^x_{i i-y}$ there are $2^{L^2}$ states.  These states can be described as the set of all string states with strings ending on the sites specified by $p$.  They can be constructed by starting with a particular state with the right string endings and acting with plaquette operators to generate all other states.  Each state in the sum has an equal weight because of the tensionless nature of the strings.

Imposing the constraint is simple in the $p$ basis.  We simply replace $|B\rangle | G \rangle$ with $$ \sum_p c_p |B_p \rangle | G_p \rangle $$ where $|G_p \rangle $ are the states described above.

Now we must ask about the spatially partitioned entanglement entropy.  In fact, we will focus on the second Renyi entropy.  Partition the system into two regions $R_1$ and $R_2$ in the $L \rightarrow \infty$ limit.  To be concrete we can take region $R_1$ to be a disk and region $R_2$ is the rest of the universe.  We will compute $S_2(R_1)$.

\subsubsection{Bosons only}
For each $|B_p\rangle $ we make a decomposition $$|B_p \rangle = \sum^{\chi_p}_{\alpha_p=1} \lambda_{\alpha_p} |B_{p \alpha_p} \rangle_1 |B_{p \alpha_p} \rangle_2$$ Just to be clear, we have a different set of $\lambda$'s for every $p$, but in the interest of avoiding a notation explosion, we try to suppress this fact. The subscript on $\alpha_p$ indicates the value of $p$.  We may take the $\lambda$'s to be real.  The full density matrix is pure $\rho = |B \rangle \langle B |$.  The reduced density matrix for region $1$ is
$$\rho_1 = \sum_{pq} c_p c_q^* \sum_{\alpha_p \alpha_q} \lambda_{\alpha_p} \lambda_{\alpha_q} |B_{p \alpha_p} \rangle \langle B_{q \alpha_q} |_1  \langle B_{p \alpha_p}|B_{q \alpha_q} \rangle_2 $$  In general, the only thing we can say about $\langle B_{p \alpha_p}|B_{q \alpha_q} \rangle_2$ is that it contains a factor of $\delta_{p_2 q_2}$ where $p_2$ and $q_2$ indicate the parts of $p$ and $q$ specifying information in $R_2$.

\begin{widetext}
We then need $\mbox{tr}_1 (\rho_1^2)$ to compute $S_2(R_1)$.  We find \begin{eqnarray} \mbox{tr}_1 (\rho_1^2) = \sum_{p q r s} c_p c_q^* c_r c_s^* \nonumber \\ \times \sum_{\alpha_p \alpha_q \alpha_r \alpha_s} \lambda_{\alpha_p} \lambda_{\alpha_q} \lambda_{\alpha_r} \lambda_{\alpha_s} \langle B_{p \alpha_p}|B_{q \alpha_q} \rangle_2 \langle B_{r \alpha_p}|B_{s \alpha_q} \rangle_2 \langle B_{q \alpha_q}|B_{r \alpha_r} \rangle_1 \langle B_{s \alpha_s}|B_{p \alpha_p} \rangle_1 \end{eqnarray}  Again, all we can say about this expression is that $\langle B_{p \alpha_p}|B_{q \alpha_q} \rangle_2$ contains a factor of $\delta_{p_2 q_2}$ and $\langle B_{p \alpha_p}|B_{q \alpha_q} \rangle_1$ contains a factor of $\delta_{p_1 q_1}$ where $p$ and $q$ could be any of $p,q,r,s$.  Note the usual replica branched surface kind of structure present in this otherwise messy looking formula: $p$ goes with $q$ and $r$ goes with $s$ in $R_2$ while $q$ goes with $r$ and $s$ goes with $p$ in $R_1$.
\end{widetext}

\begin{figure}
\includegraphics[width=.4\textwidth]{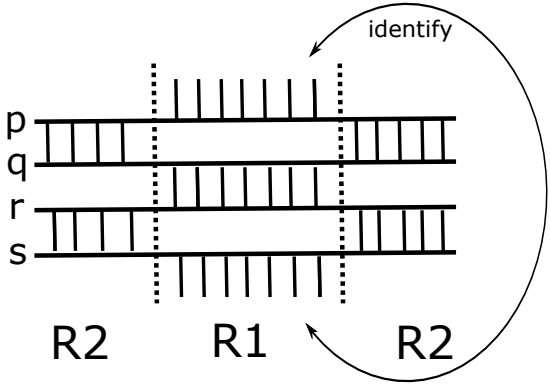}
\caption{Schematic of the computation of $\mbox{tr}(\rho^2_1)$ illustrating the different connectivities in $R_1$ and $R_2$.  Vertical bonds represent contractions of physical degrees of freedom.}
\end{figure}

\subsubsection{Bosons + gauge fields}
We now need a similar decomposition for the gauge field states.  Write $$|G_p \rangle = \sum_{I=1}^{2^{|\partial R_1| -1 }} \frac{1}{\sqrt{2^{|\partial R_1| -1 }}} |G_{p I}\rangle_1 |G_{p I}\rangle_2$$  There are important simplifications in this case.  First, the Schmidt rank is the same for all $p$.  It is always $2^{|\partial R_1| -1 }$ because we may specify the presence or absence of a string on each link piercing the boundary except for an overall global constraint requiring the number of strings entering to be even or odd depending on the number of charges inside $R_1$ (as determined by $p$).  Similarly, we may imagine for every $p$ that $A$ just labels precisely these boundary string configurations.  Thus given two choices $p$ and $q$ that agree on $R_2$ (but may disagree on $R_1$), the states from $R_2$ appearing in the decomposition above are actually the same.  In other words, we have $\langle G_{p I} | G_{q J} \rangle_2 = \delta_{I J} \delta_{p_2 q_2}$.  This is a special property of the gauge fields in this theory that permits progress to be made.

The full boson/gauge reduced density matrix is thus
\begin{eqnarray} \rho_1 = \sum_{pq} c_p c_q^* \sum_{\alpha_p \alpha_q IJ} \lambda_{\alpha_p} \lambda_{\alpha_q} |B_{p \alpha_p} \rangle \langle B_{q \alpha_q} |_1  \langle B_{p \alpha_p}|B_{q \alpha_q} \rangle_2 \\ \times \frac{1}{2^{|\partial R_1| -1 }} |G_{p I} \rangle \langle G_{q J} |_1 \langle G_{p I} | G_{q J} \rangle_2  \nonumber\end{eqnarray}

\begin{widetext}
Now again we compute $\mbox{tr}_1 (\rho_1^2)$
\begin{eqnarray} \mbox{tr}_1 (\rho_1^2) = \sum_{p q r s} c_p c_q^* c_r c_s^* \sum_{\alpha_p \alpha_q \alpha_r \alpha_s IJKL} \lambda_{\alpha_p} \lambda_{\alpha_q} \lambda_{\alpha_r} \lambda_{\alpha_s} \langle B_{p \alpha_p}|B_{q \alpha_q} \rangle_2 \langle B_{r \alpha_p}|B_{s \alpha_q} \rangle_2 \langle B_{q \alpha_q}|B_{r \alpha_r} \rangle_1 \langle B_{s \alpha_s}|B_{p \alpha_p} \rangle_1 \\ \times \left(\frac{1}{2^{|\partial R_1| -1 }}\right)^2 \langle G_{p I} | G_{q J} \rangle_2 \langle G_{r K} | G_{s L} \rangle_2 \langle G_{q J} | G_{r K} \rangle_1 \langle G_{s L} | G_{p I} \rangle_1  \nonumber \end{eqnarray}
Now we use the fact that $\langle G_{p I} | G_{q J} \rangle_2 = \delta_{I J} \delta_{p_2 q_2}$ and similarly for inner products in $R_1$.
\begin{eqnarray} \mbox{tr}_1 (\rho_1^2) = \sum_{p q r s} c_p c_q^* c_r c_s^* \sum_{\alpha_p \alpha_q \alpha_r \alpha_s IJKL} \lambda_{\alpha_p} \lambda_{\alpha_q} \lambda_{\alpha_r} \lambda_{\alpha_s} \langle B_{p \alpha_p}|B_{q \alpha_q} \rangle_2 \langle B_{r \alpha_p}|B_{s \alpha_q} \rangle_2 \langle B_{q \alpha_q}|B_{r \alpha_r} \rangle_1 \langle B_{s \alpha_s}|B_{p \alpha_p} \rangle_1 \\ \times \left(\frac{1}{2^{|\partial R_1| -1 }}\right)^2 \delta_{I J} \delta_{p_2 q_2} \delta_{K L} \delta_{r_2 s_2} \delta_{J K} \delta_{q_1 r_1} \delta_{L I} \delta_{s_1 p_1} \nonumber \end{eqnarray}  The sums over $IJKL$ now reduce to a sum over $I = J = K = L$ giving a factor of $2^{|\partial R_1 | -1}$.  Moreover, the explicit delta functions may be absorbed into the corresponding boson overlaps e.g. $ \langle B_{p \alpha_p}|B_{q \alpha_q} \rangle_2  \delta_{p_2 q_2} = \langle B_{p \alpha_p}|B_{q \alpha_q} \rangle_2 $ since $  (\delta_{p_2 q_2} )^2 =  \delta_{p_2 q_2} $ and $\langle B_{p \alpha_p}|B_{q \alpha_q} \rangle_2 $ already contains such a factor.  Thus we find
\begin{eqnarray}\mbox{tr}_1 (\rho_1^2) = \sum_{p q r s} c_p c_q^* c_r c_s^* \sum_{\alpha_p \alpha_q \alpha_r \alpha_s} \lambda_{\alpha_p} \lambda_{\alpha_q} \lambda_{\alpha_r} \lambda_{\alpha_s} \langle B_{p \alpha_p}|B_{q \alpha_q} \rangle_2 \langle B_{r \alpha_p}|B_{s \alpha_q} \rangle_2 \langle B_{q \alpha_q}|B_{r \alpha_r} \rangle_1 \langle B_{s \alpha_s}|B_{p \alpha_p} \rangle_1 \left(\frac{1}{2^{|\partial R_1| -1 }} \right) \\ = \left(\frac{1}{2^{|\partial R_1| -1 }} \right) \mbox{tr}_1(\rho^2_1)_{\mbox{bosons only}} \nonumber\end{eqnarray}\end{widetext}

This final line implies that $S_2(R_1) = S_2(R_1)_{\mbox{bosons only}} + (|\partial R_1| - 1) \ln{2}$.  In fact, this relation holds for all Renyi entropies and hence holds for the entanglement entropy as well.  Note that in this case the gauge field contribution is in fact independent of $n$.  Thus if we parameterize the entanglement entropy at the critical point as $S = \alpha L - \beta$ then we have $S_{XY} = \alpha L - \beta_{XY}$, $S_{\mathbb{Z}_2} = \alpha' L - \beta_{\mathbb{Z}_2}$, and $S_{XY^*} = \alpha'' L - \beta_{XY^*}$ with $\beta_{XY^*} = \beta_{XY} + \beta_{\mathbb{Z}_2}$.  We note that $\beta_{Z_2}$ does not depend on region shape while $\beta_{XY}$ will (being associated with a critical point), hence there is a precise distinction between these two contributions to $\beta_{XY^*}$.

\subsection{More complicated regions}
We have just seen via an explicit wavefunction computation that at the critical point the entanglement entropy has the form $\alpha L - \beta$ with $\beta = \beta_{XY} + \beta{Z_2}$.  However, we only established this result for a simply connected region.  The alternative derivation to be presented below based on a partition function argument also only tells us the answer for a simple disk geometry.  The question thus remains, how does $\beta$ depend on region shape and topology?  As we will argue in a later section, the gapless contribution to $\beta$ has both non-trivial shape dependence as well as non-trivial topological dependence.  On the other hand, the topological contribution, being associated with local constraints, depends only on the topology of the region.

In a gapped phase of matter in two dimensions, the entanglement entropy of a region $R$ of linear size $L$ takes the form $S = \alpha L - \beta(R) = \alpha L - \beta(\mbox{disk}) N_{\partial}$ where $N_{\partial}$ is the number of disconnected boundaries of $R$.  This formula is valid up to exponentially small corrections in $L/\xi$.  Corners will not contribute logarithmically growing contributions for $L \gg \xi$, however they can contribute terms of the $\ln{\xi/\epsilon}$ which must be separated out from the smooth part $\beta$.  If we can arrange, in numerical computations, say, to study regions without corners, then $\beta$ can be extracted from finite size scaling of $S(L)/L$.  If corners cannot be avoided, as is often the case on a lattice, then extra care must be taken to remove the corner contributions.

\begin{figure}
\begin{center}
\includegraphics[width=.4\textwidth]{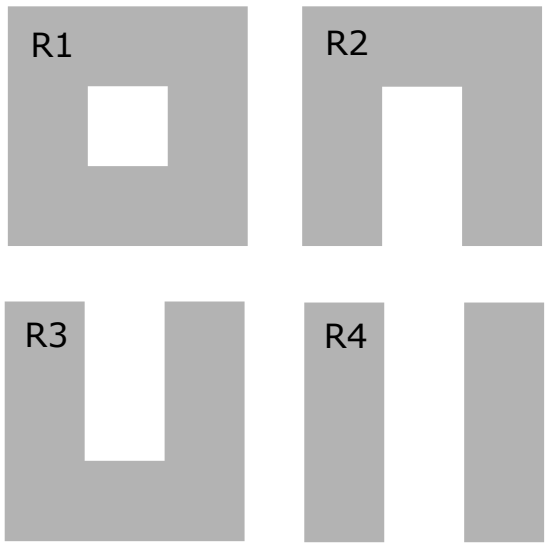}
\end{center}
\caption{Four regions defining the Levin-Wen procedure for extracting the topological entanglement entropy.  The regions are chosen so that all boundary and corner terms will cancel.}
\end{figure}

Protocols to achieve this separation for gapped phases were presented in Refs. \onlinecite{topent1,topent2}.  In the case of Ref. \onlinecite{topent2}, the entanglement entropy of four regions are subtracted in a certain way to remove all boundary law and corner terms.  The four regions, shown in Fig. 2, call them $R_i$ $(i=1,...,4)$, look like an annulus ($R_1$), two half moons ($R_2$ and $R_3$), and two disconnected strips $R_4$.  All regions are topologically collections of disks except for the annulus $R_1$.  The prescription is then $(S(R_1) - S(R_2)) - (S(R_3) - S(R_4)) = (- 2 \beta + \beta) - (- \beta + 2 \beta) = - 2 \beta$.

We may also ask, in the context of $Z_2$ gauge theory, say, what happens when either the electric or magnetic charges are allowed to fluctuate.  This question may be given a simple answer in the context of the toric code by putting one of the constraint terms to zero.  There is now a macroscopically degenerate space of ground states, but we may still extract a ``topological entropy" via the same procedure (volume, area, and corner terms are all subtracted out).  The important point here is that the constant term no longer depends on the number of disconnected components in a simple way.  For example, two disconnected strips still give $- 2 \beta = - 2 \ln{2}$, but the annulus now only gives $- \beta = - \ln{2}$.  The prescription now gives $- \beta$, half the original result.  The interpretation of this result is that half the constraint structure has been destroyed with the other half remaining intact.

Now we ask what happens in a gapless state.  In the case of interest, electric charges are allowed to fluctuate, but their motion is not completely washed out (no large number of degenerate states).  First, $\beta$ will now depend on region shape in a non-trivial way.  For example, a long thin strip (with smoothed corners) of length $L$ and width $w$ would have $\beta \sim L/w$.  Also, corners will now give a logarithmic contribution of the form $f(\theta) \ln{(L/\epsilon)}$ with $\theta $ measuring the deviation from a straight line and $f$ an even function.  We also note that these considerations apply for regions embedded into an otherwise infinite system.  Finite size corrections will modify all these statements, and some considerations along these lines will be relevant when comparing to numerical data.  Already the prescriptions in Refs. \onlinecite{topent1,topent2} are less appealing because, although they subtract the corner dependence in a sensible way, the shape dependent parts of $\beta$ are no longer guaranteed to nicely cancel.  We return to this point later.

\begin{figure}
\begin{center}
\includegraphics[width=.4\textwidth]{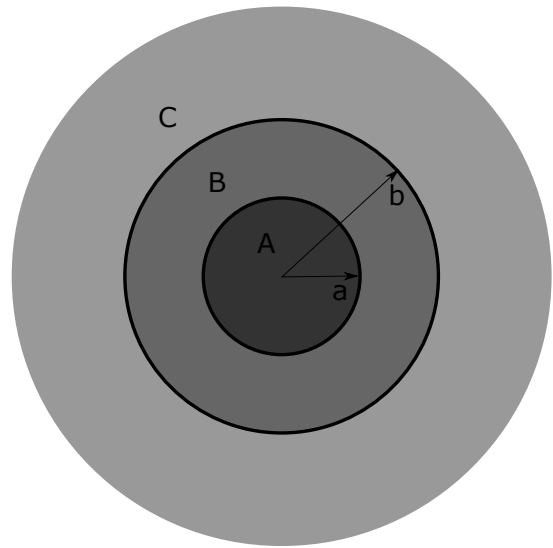}
\end{center}
\caption{A smoothed version of the annulus entering the Levin-Wen protocol.  We need the structure of the reduced density matrix of this geometry as a function of the radii $a$ and $b$. }
\end{figure}

For now, we want to argue that the topological part of $\beta$ for the annulus is $ 2 \ln{2}$ even at the critical point.  Such a result is quite different from the answer obtained by putting one of the constraint terms in the toric code to zero.  Since we are no longer dealing with a gapped phase, we must specify the geometry of the annulus more carefully.  Let $a$ be the inner radius and $b$ the outer radius.  We consider a limit of $a,b \rightarrow \infty$ with $a/b \rightarrow 0$ e.g. the outer radius is going to infinity faster than the inner radius.  Call $A$ the inner disk, $B$ the annulus, and $C$ the rest of the system as shown in Fig. 3. Because the state of the whole system $ABC$ is pure, the entanglement entropy $S(B)$ of the annulus is equal to $S(AC)$.  The density matrix for this composite system is expected to factorize in the limit we consider giving $S(AC) \approx S(A) + S(C)$.  To quantify this expectation, we turn to the mutual information $\mathcal{I}(A,C) = S(A) + S(C) - S(AC)$.  The mutual information bounds connected correlations according to $\langle \mathcal{O}_A \mathcal{O}_C \rangle_c^2 \leq ||\mathcal{O}_A||^2 ||\mathcal{O}_C||^2 \mathcal{I}(A,C)$ as shown in Ref. \onlinecite{minfo}, and although we do not prove it, we expect this bound to be tight at the critical point. \footnote{One situation where the mutual information does not decay appears when we consider ``cat states" in symmetry breaking phases.  Such states can be relevant when considering finite size numerical simulations.}   Thus the mutual information decays as fast as square of the slowest decaying connected correlation function of normalized gauge-invariant operators in $A$ and $C$.  This means the density matrix $\rho_{AC}$ factorizes as $a,b \rightarrow \infty$ as a power law in the separation $b-a$.  Using this information, we can conclude that the shape dependent part of $\beta$ is simply that of two disks, up to power law corrections.  This also suggests that the topological parts is also equivalent to that of two disks, but given the importance of this claim, we further justify it.

Now we turn to the topological part.  Let us imagine fixing, as in the wavefunction calculation above, the distribution of $Z_2$ charges inside the annulus.  Suppose there are an even number of charges.  Without gapless charges, we would be forced to say that an even number of electric strings enter from both $A$ and $C$, a situation we call $(0,0)$. However, with fluctuating charges we could also have an odd number of charges in $A$ and $C$.  Now there will be an odd number of strings entering from both $A$ and $C$, but still an even number of strings entering the annulus through all its boundaries taken together.  This situation we call $(1,1)$.  In the toric code with constraint removed, the existence of these two sectors reduces $\beta$ from $ 2 \ln{2}$ to $\ln{2}$.  The important question is, does the state in region $B$ care if we have an even or odd number of particles in $A$ and $C$.  Is the boson wavefunction (before the strings are attached) more like \textit{(i)} $(|(0,0)\rangle_{AC} + |(1,1) \rangle_{AC}) | \mbox{even} \rangle_B$ or \textit{(ii)} $(|(0,0)\rangle_{AC} | \mbox{even},(0,0)\rangle_B + |(1,1) \rangle_{AC} | \mbox{even}, (1,1) \rangle_B)$. Case \textit{(i)} is typical of the superfluid phase where bosons may fluctuate freely.  In such a phase the the correlator $\langle b_A b^+_C \rangle $, containing an operator that moves us from $(0,0)$ to $(1,1)$, is finite because the state is long range ordered.  This correlator is also finite in the cartoon state \textit{(i)}.  Case \textit{(ii)} is typical of the Mott insulating phase where the bosons are locked in place.  In such a phase the correlator $\langle b_A b^+_C \rangle $ is zero, as it is in the cartoon state \textit{(ii)} (assuming the states in $B$ are orthogonal).  Because at the critical point the correlator $\langle b_A b^+_C \rangle $ decays with a universal power law, we argue that the critical point is more like the Mott insulator than the superfluid phase.  Hence we should find that the topological part of $\beta$ is $2 \ln{2}$.  We emphasize again that this is all true up to power law corrections at the critical point.

We briefly comment on the situation at finite temperature to illustrate the scales involved.  We imagine starting at $T=0$ in the Mott insulating phase, then moving up to encounter first the quantum critical region, then a higher temperature region where the boson charge dynamics are disordered, and finally to an even higher temperature region beyond vortex gap.  The topological entropy, defined by the usual protocol, should be $- 2 \ln 2$ in the insulating phase and in the quantum critical region as we just argued.  Above the dynamical scale set by the boson motion, we should find $-\ln{2}$ as is appropriate for the electrically disordered toric code.  Finally, the second constraint is lifted above the vison gap, and we find no topological entropy.

\section{Partition function argument}
\subsection{Conformal transformation}
Now we give a more general argument for the structure of the entanglement entropy by mapping the entanglement entropy of a disk onto a problem of computing the partition function of a sphere, see Ref. \onlinecite{holoee_deriv}.  This transformation works whenever the theory in question is a conformal field theory.  This includes topological field theories (a highly degenerate case of conformal field theory) and most $z=1$ scale invariant theories.  However, we should be careful applying these results to $z \neq 1$ scale invariant theories.

We want to compute the entanglement entropy of a disk $B^2$ in $d=2$ dimensions, or more generally, a ball $B^d$ of radius $R$ in $d$ spatial dimensions.  We begin by writing $d+1$ dimensional Minkowski space metric in spherical coordinates as
\begin{equation}
ds^2 = - dt^2 + dr^2 + r^2 d\Omega_{d-1}^2,
\end{equation}
where $d\Omega_{d-1}^2$ is the unit round metric on the $d-1$ sphere.  In the case of $d=2$, we have $d\Omega_{2-1}^2 = d\phi^2$ with $\phi \sim \phi + 2 \pi$.  Now consider the coordinate transformation given by
\begin{eqnarray}
t = R \frac{\cos{\theta} \sinh{\tau/R}}{1 + \cos{\theta} \cosh{\tau/R}} \nonumber \\
r = R \frac{\sin{\theta}}{1 + \cos{\theta} \cosh{\tau/R}}.
\end{eqnarray}
The metric in these new coordinate now reads
\begin{eqnarray}
\label{desitter}
ds^2 = F^2 \left( - \cos^2{\theta} d\tau^2 + R^2 (d\theta^2 + \sin^2{\theta} d\Omega^2_{d-1})\right) \nonumber \\
F = (1 + \cos{\theta} \cosh{\tau/R})^{-1}.
\end{eqnarray}
The conformal factor $F^2$ may now be removed by a conformal transformation.

In these new coordinates the ``entanglement Hamiltonian" is nothing but the generator of time translations for the variable $\tau$ as shown in Ref. \onlinecite{holoee_deriv}.  In other words, $\rho_{B_d} = Z^{-1} \exp{(- 2 \pi R H_{\tau})}$ with $2 \pi R$ playing the role of the inverse temperature.  If we ignore the conformal factor, as we are allowed to do for a conformal field theory, then we may compute the partition function $Z = \mbox{tr}(\exp{(- 2 \pi R H_{\tau})})$ by continuing the metric above to imaginary time $\tau = - i \tau_E$ and evaluating the partition function of the conformal field theory on the resulting space.  Remarkably, there is a simplification, because when we consider the Euclidean version of \ref{desitter} it can be shown to give the round metric on $S^{d+1}$.  Thus, to obtain the entanglement entropy of a ball of radius $R$, we need the conformal field theory partition function on a sphere of radius $R$ (for details see Ref. \onlinecite{holoee_deriv}).  We also note that in principle we have access to the full entanglement Hamiltonian.

The simplest calculation that illustrates this tool is a computation of the entanglement entropy of a disk in topological liquids in $2+1$ dimensions. Topological field theories are certainly conformal field theories since they do not care about the metric of spacetime at all.  To compute $Z(S^3)$ we begin with the fact that $Z(S^2\times S^1) = 1$.  This is the statement that a topological phase has a unique ground state on the sphere.  The spacetime $S^2\times S^1$ may be cut open along the $S^2$ to yield two copies of the solid torus $B^2 \times S^1$.  Since $\partial B^2 \times S^1 = S^1 \times S^1$, $Z(B^2\times S^1) = |\Psi \rangle$ is a state in the Hilbert space of the torus generated by imaginary time evolution.  This state is normalized since $1 = Z(S^2\times S^1) = \langle \Psi | \Psi \rangle$.  Instead of gluing the tori back together directly, we first make an $S$ modular transformation of one of the boundary tori which exchanges the two non-contractible surface loops.  Now gluing the tori together yields $S^3$ as shown in Fig. 4. The modular transformation is implemented using the modular $S$-matrix $\mathcal{S}^a_b$, and a simple calculation gives $Z(S^3) = \langle \Psi | \mathcal{S} |\Psi \rangle = \mathcal{S}^0_0$.  Since $S^0_0 = 1/\mathcal{D}$ and using the fact that the ordinary Hamiltonian of a topological phase is zero, we conclude that $S(B^2) = \ln{Z(S^3)} = - \ln{\mathcal{D}}$ as originally shown in Refs. \onlinecite{topent1,topent2}.  In fact, using the exponentially fast factorization of the density matrix for widely separated regions and the topological invariance of $\beta_{\mbox{topo}}$, we may immediately establish that $\beta_{\mbox{topo}} = - N_{\partial} \ln{\mathcal{D}}$ where $N_{\partial}$ is the number of boundaries of the region of interest.

\begin{figure}
\begin{center}
\includegraphics[width=.46\textwidth]{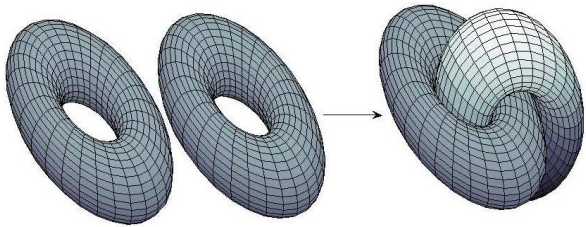}
\end{center}
\caption{Visualizing the modular transformation that maps two copies of the solid torus $B^2 \times S^1$ to the pair of interlocking solid tori on the right.  Expanding the interlocking tori to fill all of space and gluing them along their mutual boundary produces $S^3$.}
\end{figure}

\subsection{General computation}
Using the tools we just introduced, we now give another heuristic argument for the conclusion that $\beta_{XY^*} = \beta_{XY} + \beta_{\mathbb{Z}_2}$.  Since the $XY^*$ critical point is a conformal field theory, the entanglement entropy of a disk may be obtained from $Z_{XY^*}(S^3)$.  We take $Z_{XY}(S^3)$ and $Z_{\mathbb{Z}_2}(S^3)$ as given.  To obtain $XY^*$ from $XY\otimes \mathbb{Z}_2$, we must bind the two sectors together by attaching $\mathbb{Z}_2$ Wilson lines to the bosons.  However, in the extreme topological limit we considered above, all Wilson lines are tensionless objects in the gauge sector.  Thus assuming a large gap to visons, the attachment of Wilson lines to bosons leaves the boson dynamics essentially unaffected as they acquire neither interesting phases nor energetic penalties.  For example, the projected wavefunction we wrote down above has the same boson density correlators as the unprojected wavefunction.  Hence we argue that for the purposes of universal physics we have $Z_{XY^*}(S^3) \approx Z_{XY}(S^3) Z_{\mathbb{Z}_2}(S^3)$.  In this topological limit we have $Z_{\mathbb{Z}_2}(S^3) = 1/\mathcal{D} = 1/2$, and we recover the fact that $\beta_{XY^*} = \beta_{XY} + \beta_{\mathbb{Z}_2}$.

The advantage of this argument is that it is simple and can be easily generalized.  Consider any other bosonic critical point between an insulating and a broken symmetry phase where the bosons are charged under some discrete group $G$.  By gauging the group $G$ we can describe deconfined transitions between exotic insulators and conventional broken symmetry phases.  The projection procedure works as above, namely, we attach Wilson lines to the bosons according to the  representation of $G$ they carry.  However, even when $G$ is non-Abelian, these pure charge Wilson lines (chargons) have trivial braiding properties.  Furthermore, assuming as before that the magnetic excitations (fluxons and flux-charge composites) are gapped, we can conclude that the boson dynamics are not seriously modified by the projection.  This allows us to conclude that $Z_{\mbox{gauged bosons}} \approx Z_{\mbox{bosons}}Z_G$ which, via our argument above, gives $\beta_{\mbox{gauged bosons}} = \beta_{\mbox{bosons}} + \beta_G$.

We wish to emphasize at this point that the partition function argument applies only the special case of the disk.  Because of the topological nature of $\beta_G$, we expect that the disk result will hold true for any other simply connected region (or multiple such regions).  We will simply need to take into account the shape dependence of $\beta_{\mbox{gapless}}$.  However, we cannot directly probe the topological contribution to more complicated geometries such as the annulus using this argument.  Nevertheless, the general considerations laid out above for the special case of the $XY^*$ critical point also lead us to conjecture that the topological part of the entanglement entropy for the annulus remains at its value inside the topological phase.  To the extent that the annulus entropy is equal to that of two factorized disks, the partition function argument above rigorously establishes our claim.

\section{Non-compact $CP^1$ model}
\subsection{Description of transition}
Here we consider certain square lattice spin-$1/2$ quantum antiferromagnets realizing a direct continuous transition between a Neel ordered state and a valence bond solid (VBS) state.  Since these two phases break different symmetries present in the microscopic Hamiltonian, a continuous transition between them is non-generic in the Landau theory.  Unless we fine tuned the model, we would expect either a first order transition or coexistence.  Thus the critical point, if there is one, cannot be the conventional $O(3)$ critical point.  The mechanism for the phase transition is actually the condensation of spinons, bosons with fractional spin which are confined in the Neel and VBS phases, but which are liberated at the critical point.  The low energy description of the transition is given in terms of the so-called non-compact $CP^1$ model.  This model consists of bosonic spinons carrying fractional charge coupled to a fluctuating gauge field.  Right at the critical point, the instantons which typically confine a $U(1)$ gauge theory in $2+1$ dimensions become irrelevant.  Lattice models that realize this transition differ from the pure field theory by irrelevant operators that flow to zero in the low energy limit.

\subsection{Irrelevant operators}
Having described the basic physics of the Neel to VBS transition and its potential description in terms of the non-compact $CP^1$ model, we would like to answer the following question.  Given that monopoles are irrelevant at the critical point, so that the gauge field is effectively non-compact, will a field theory computation, say using continuum QED, give the correct universal features of the entanglement entropy?  For example, will $\beta$ as defined by the lattice model at its critical point be the same as $\beta$ computed using the field theory.  We argue that this is so since these two models differ only by irrelevant operators at the critical point.  In particular, the microscopically allowed instanton operators are all irrelevant at the critical point.  Of course, away from the critical point there will be a complicated crossover description, for example, the Neel phase will have its own value of $\beta$ due to gapless spin waves.

Another way to see this is as follows.  We can always formulate the transition in terms of the VBS order parameter so long as we include the appropriate WZW terms responsible for adding quantum numbers to the vortices of the VBS order parameter (see Refs. \onlinecite{PhysRevLett.95.036402,PhysRevB.74.064405}).  The presence of monopoles is related to the four fold anistropy of the VBS order parameter.  Since this anistropy is irrelevant at the critical point, we may as well study the model with full $SO(2)$ symmetry in the presence of the WZW term.  This model may again be shown to be equivalent to the non-compact gauge theory via a duality transformation.

Note that this is not conflict with the comment made in the introduction about the importance of remembering the compactness of order parameter fields.  For example, we should not expect $\beta$ in the superfluid phase to agree with $\beta$ of a free boson, as one is compact and the other is not.  The essential point is that the compact model may be viewed as a gauged version of the non-compact model, and this gauging procedure means that many operators that are apparently sensible in the low energy theory are in fact not allowed.  Related to this, there must also be new operators, vortex operators, that correspond to the topological defects of the model.  In the case of the gauge description relevant for the Neel to VBS transition, the low energy theory already contains all the fields needed to insure that no unphysical operators can be constructed.  Our claim is thus that the compactness of a $U(1)$ gauge field can be forgotten when monopoles are irrelevant even though the compactness of an order parameter cannot.

\subsection{Renormalization group flow}
Besides arguing for the insensitivity of the universal parts of the entanglement entropy on irrelevant operators, we would also like to comment on the extent to which ``exotic" critical points may be regarded as more entangled that simple symmetry breaking critical points.  By analogy with topological phases in two dimensions, we might suggest that a larger $\beta$ corresponds to more entangled critical point.  Certainly it seems that phases with greater long range entanglement and more complicated topological order generally have a larger total quantum dimension.  However, we must also be careful to note that we can typically increase $\beta$ by simply tensoring together many weakly interacting degrees of freedom.  The most sensible comparisons of $\beta$ between critical points would thus be made when those critical points describe transitions out of the same phase, so that we in some sense fix the number of degrees of freedom.

For example, we could consider two transitions out of the Neel ordered state.  The first is a conventional symmetry breaking transition while the second is the deconfined critical point we have been considering.  If $\beta$ were larger for the deconfined critical point, then we would be justified in saying that it was more entangled than the conventional $O(3)$ critical point.  Note that we might in general have to be careful about choosing the right region when trying to establish such a claim.  Below we compare $\beta$s for disks.

In fact, a conjecture is available that would imply the deconfined critical point is more entangled, suitably defined, than the conventional $O(3)$ critical point.  In Ref. \onlinecite{highercurv_ee} it is proposed that a certain subleading term in the entanglement entropy of half of a sphere is monotonic under RG flow.  This claim is supported with a variety of holographic calculations, and the claim in $1+1$ and $3+1$ dimensions appears to be established thanks to c-theorems in those dimensions, see Ref. \onlinecite{4dctheorem}.  Assuming this claim is true in $2+1$ dimensions, we can say that the deconfined critical point is more entangled than the $O(3)$ critical point because it can flow to the usual $O(3)$ point under relevant deformations \footnote{For example, the single instanton operator is relevant at the non-compact $CP^1$ critical point with the RG flow terminating at the $O(3)$ critical point.}.  To be precise, if the full system is tuned to criticality on a sphere, then the entanglement entropy of half the sphere contains a subleading term that decreases under RG flow from the deconfined critical point to the $O(3)$ critical point.  We must be careful in extending these claims too far, however, since the results of Refs. \onlinecite{lif_ee1,lif_ee2,lif_ee3} show that $\beta$ can be negative for some Lifshitz critical points.  It is not clear how to reconcile this behavior with a general entanglement c-theorem.

\section{Extracting universal terms and numerics}
\subsection{General considerations}
We now return to the somewhat subtle issue of properly defining and extracting the universal coefficient $\beta$.  In practice, the simplest method is simply to compute $S(L)$ for a variety of region sizes and fit to $S(L) = \alpha L - \beta + \mathcal{O}(1/L)$.  This method seems to work well enough, but it is dangerous because of the possibility of mixing some of the non-universal component in $\beta$.  This is especially problematic when the region is question has corners, as it often does in a lattice formulation.  Corners are expected to give a contribution like $\log{(L/\epsilon)}$ and thus changes in $\epsilon$, the UV regulator, can look like changes in the ``universal" parameter $\beta$.  It is therefore important to have a well defined way, at least in principle, to extract $\beta$.

For topological phases in two dimensions the constant $\beta$ is related to the total quantum dimension of the topological phase.  It can be measured independent of any non-universal components of the entanglement entropy using a particular set of subtractions that we have already discussed.  This kind of procedure is familiar from the theory of mutual information $\mathcal{I}(A,B) = S(A) + S(B) - S(AB)$ where the subtractions can be motivated in the continuum by the desire to remove the non-universal boundary law terms (singularities still arise if $A$ and $B$ touch, see Ref. \onlinecite{qft_mi}).  The finite correlation length in a gapped phase both implies that distant parts of the boundary may be treated separately and that corners do not give special contributions.  Both of these statements may fail in a gapless phase.

As already mentioned, corners are known to give a logarithmic correction to the entanglement entropy.  This correction is analagous to the singularities that appear in Wilson lines with sharp corners in gauge theory.  From the point of view of the renormalization group argument sketched in the introduction, the logarithm arises because the corner contributes order one entanglement at every RG step.  Thus if we integrate the corner contribution from $\epsilon$ to $L$ we find a contribution to the entanglement entropy going like $\int dr/r \sim \log{(L/\epsilon)}$.  However, one can check that the substraction used to extract $\beta$ for topological phases also correctly removes any potential corner terms.  This works in part because the coefficient of the logarithmic term is an even function of the angle measuring the deviation from straightness i.e. a $\pi/2$ corner gives the same contribution as a $3 \pi / 2$ corner.  On the practical side, the contributions from corners can often be quite small.  It can be further mitigated by considering very slight corners, however, such an arrangement often requires a very complicated region shape with many small corners.  It is possible that the best proposal is to simply use a ``rough" circle on the lattice with many small lattice scale corners trusting the RG flow to quickly smooth out these irregularities.

The other potential concern is the slow (power law) fall-off of correlations in a gapless phase.  For example, the subtraction to extract $\beta$ relies on the assumption that the subleading term in the entanglement entropy of an annulus is $- 2\beta_{\mbox{disk}}$.  However, we can already see a problem with this assumption, namely that $\beta$ can have non-trivial shape dependence, so we must ask if the annulus is $- 2\beta_{\mbox{disk}}$ or something else entirely.  It is clear that for widely separated regions the entanglement entropies will add, however the annulus geometry is not exactly of this type.  We may gain some intuition for this situation by appealing to the twist operator formalism for entanglement entropy developed in Ref. \onlinecite{qft_mi}.  That formalism suggests the following test: if the correlation functions of twist fields decay quickly enough then the inner and outer ring of the annulus may be treated separately (up to power law decaying corrections).  To estimate their decay, we use the fact that at large separation the mutual information (which involves integrating the twist field correlator over distant region boundaries) bounds the connected two point function of local operators.  If $\Delta_0$ is the smallest scaling dimension among all local gauge-invariant operators, then we expect $\mathcal{I}(x) \sim 1/x^{4 \Delta_0}$ at large separation $x$.  The unitarity bound in $2+1$ dimensional conformal field theories requires that $\Delta_0 \geq 1/2$ for scalar operators.  Practically speaking, the corrections could be quite large for a finite annulus, so one would have to study a range of annuli sizes to extract the constant term from finite size scaling.

All these considerations lead to a much more complicated situation in the case of gapless phases.  However, we are still interested in studying the entanglement properties of critical points in terms of $\beta$ and corner contributions.  The simplest protocol is thus to use smooth simply connected regions to extract $\beta$ from finite size scaling.  One could further study the shape dependence of $\beta$ as a diagnostic of our proposed decomposition into gapless and topological contributions (although this decomposition may not hold in more complicated theories where the gapless states are not pure charge).  Unfortunately, this may be nearly impossible in a lattice setup where corners are almost inevitable.  We could still isolate the shape dependent part of $\beta$ by studying, for example, $S_{\mbox{square}} - S_{\mbox{rectangle}}$.  This subtraction will approximately remove corner terms, and if we keep the perimeter of the rectangle equal to that of the square, the subtraction will depend only on the aspect ratio of the rectangle.  However, this is still not fully satisfactory since the topological part of $\beta$ will simply be subtracted away.  Another possibility is to consider the system on a torus and study regions that wrap around the loops of the torus.  In this way it is possible to have a region on a lattice without corners, but at the cost of exposing the entanglement entropy to additional topological subtleties.  However, at least in the case of gapped topological phases, these subtleties can be understood and taken into account.

Let us summarize the situation as we see it.  If we do use the usual topological protocol for gapless systems, then we will obtain the topological part plus a sum of terms that depend on region shape.  It is possible that these shape dependent terms can be calculated or fit numerically and subtracted away.  The result would be the topological part of the entropy, however, we stress that this would be a quite complicated procedure in general.  Corners are present and are hard to avoid on the lattice.  Regions wrapping around the entire compact space can avoid corners but come with additional complications.  There will be power law corrections to the various asymptotic statements we have made in this paper, as well as the aforementioned shape dependent universal terms coming from the gapless modes.  Results from field theory often place special emphasis on the properties of simple spaces like the sphere and simple regions like disks, regions that are hard to replicate on the lattice.  Finally, we remind the reader again that finite size corrections can have a very profound impact on the entanglement entropy e.g. freezing out gapless degrees of freedom.  Despite all this complexity, as we have illustrated in this paper, it is possible to gain control of the universal parts of the entanglement entropy in a wide variety of geometries.  Thus we believe we have the necessary flexibility to untangle the complexity of universal gapless entanglement.

\subsection{Numerical issues}
In the numerical investigation of the $XY^*$ transition \onlinecite{melko0}, there were some puzzles associated with the observed entropies.  Those authors considered a hard core boson model with nearest neighbor hopping on the kagome lattice with an extra constraint term associated with hexagons.  The Hamiltonian is $H = - t \sum_{<rr'>}b_r^+ b_{r'} + b_{r'}^+ b_r + V \sum_{\mbox{hex}} (\sum_{i \in \mbox{hex}} n_i)^2$.  This model has a superfluid to $Z_2$ insulator transition with increasing $V$ located at $V/t \approx 7$.  In the absence of $V$ the model is in a superfluid phase where the elementary excitations are phonons.  The speed of these excitations will be set by $t$ in units of the lattice spacing since there is no other scale.  To probe the contributions to $\beta$ from critical degrees of freedom we must consider a system of large enough size and low enough temperature so that these low energy degrees of freedom can fluctuate.  At the critical point, similar considerations apply, and although the velocity can be renormalized, we still expect relativistic low energy modes with speed $\sim V \sim t$.  Given the lattice sizes, region sizes, and temperatures considered in Ref. \onlinecite{melko0}, it is possible that some of these gapless modes were frozen.  Thus we cannot be sure that the shape dependent part of the entropy was being effectively sampled.

We have also predicted a topological part at the critical point (as measured by the usual protocol) given by the full $- 2 \ln{2}$ of the insulating phase.  This is so despite the presence of fluctuating $Z_2$ charges as we argued above.  However, it is also possible, depending on the details of the temperature and the energy scale of the charge motion, that these fluctuating charges are moving incoherently, a situation that then gives a topological part of $- \ln{2}$. If we are really probing the ground state properties, then we believe the topological part should be $-2 \ln{2}$.  Even if this number is not observed exactly, the fact that topological contributions to the entropy are not continuously variable means we can still be confident we have a $Z_2$ deconfined critical point or phase.

The results of Ref. \onlinecite{melko0} are quite interesting, and more work is needed on both sides to achieve a completely satisfactory comparison between theory and experiment.  We hope that the successes of Ref. \onlinecite{melko0} lead to other calculations of the entanglement entropy at quantum critical points, conventional or otherwise.

\section{Conclusions}
We have studied the entanglement structure of a variety of deconfined quantum critical points focusing on $2+1$ dimensions.  We have computed the universal terms in the entanglement entropy in terms of the same data at conventional critical points and in topological phases.  We have also argued for the insensitivity of the universal terms to irrelevant operators in the low energy description.  Based on our explicit calculations and on a conjectured c-theorem for entanglement entropy, we can confidently assert that deconfined critical points are more entangled, in the sense discussed above, than corresponding conventional critical points.  Finally, we carefully analyzed the structure of entanglement for more complicated region geometries and discussed in detail various protocols for extracting universal terms.

Given the power of the partition function argument for computing the entanglement entropy of balls in a CFT, there are many interesting systems that could be attacked using this technology.  It would also be very interesting to investigate further the full entanglement Hamiltonian in a CFT, which is known for some special regions.  Despite the boundary law for entanglement entropy, the entanglement Hamiltonian in a CFT is not well localized to the boundary, so it would be interesting to understand more precisely how local the entanglement Hamiltonian is.  There is also considerable opportunity to carry out further numerical studies of entanglement in deconfined critical points (and, in fact, in conventional critical points).  Thanks to powerful new perspectives and tools, there is a growing body of analytical results in higher dimensions that can be compared with numerical calculations.

\textit{Acknowledgements}
BGS is supported by a Simons Fellowship through Harvard University.  TS is supported under grant number DMR-1005434.  We thank Maissam Barkeshli and Rob Myers for helpful conversations on these topics.

\bibliography{xystar_paper}

\end{document}